Title: Prognostic factors associated with success rates of posterior orthodontic miniscrew implants: a subgroup meta-analysis

Article Type: Review Article

Abstract: Objective: To systematically review previous studies and to assess the combined odds ratio (OR) regarding prognostic factors affecting the success of miniscrew implants(MIs) inserted into the buccal posterior region via a subgroup meta-analysis.
Methods: Three electronic searches were conducted to obtain articles in English limited to clinical human studies published prior to March 2015. The outcome measure was the success of MIs. Patient factors included age, gender, and jaw; the MI factors included length and diameter. A meta-analysis was then performed based on 17 individual studies. The quality of each study was assessed for non-randomized studies and quantified using the Newcastle–Ottawa Scale. The outcome of the meta-analysis was a combined OR. Subgroup and sensitivity analyses based on the study design, study quality, and sample size of MI were performed.
Results: Significantly higher success rates were revealed in the maxilla, in patients of age 20 and over, and in long length(≥8mm) and large diameter(>1.4mm). All subgroups acquired homogeneity and the combined OR of the prospective studies(OR = 3.67, 95% CI = 2.10 – 6.44) was significantly higher in the maxilla than in the retrospective studies(OR = 2.10, 95% CI = 1.60 – 2.74).
Conclusions: In terms of clinical application, if we know the success rate of each type of screw (length/diameter) as well as its region (maxilla/mandible), we will be able to reduce the failure rate based on the findings of this meta-analysis.

# Prognostic factors associated with success rates of posterior orthodontic miniscrew implants: a subgroup meta-analysis

## Abstract

**Objective**: To systematically review previous studies and to assess the combined odds ratio (OR) regarding prognostic factors affecting the success of miniscrew implants(MIs) inserted into the buccal posterior region via a subgroup meta-analysis.

**Methods**: Three electronic searches were conducted to obtain articles in English limited to clinical human studies published prior to March 2015. The outcome measure was the success of MIs. Patient factors included age, gender, and jaw; the MI factors included length and diameter. A meta-analysis was then performed based on 17 individual studies. The quality of each study was assessed for non-randomized studies and quantified using the Newcastle-Ottawa Scale. The outcome of the meta-analysis was a combined OR. Subgroup and sensitivity analyses based on the study design, study quality, and sample size of MI were performed.

**Results**: Significantly higher success rates were revealed in the maxilla, in patients of age 20 and over, and in long length(≥8mm) and large diameter(>1.4mm). All subgroups acquired homogeneity and the combined OR of the prospective studies(OR = 3.67, 95% CI = 2.10 − 6.44) was significantly higher in the maxilla than in the retrospective studies(OR = 2.10, 95% CI = 1.60 − 2.74).

**Conclusions:** Significantly higher success rates of MIs inserted in maxilla, patients of 20 years and over, and MIs with long length (≥8mm) and large diameter (>1.4mm) were shown.

# INTRODUCTION

Despite the number of studies investigating various prognostic factors affecting the success of miniscrew implants (MIs) insertion, conflicting results have made the identification of critical factors controversial. Some studies have reported no significant differences between the success rates of MIs inserted in the mandible and those inserted in the maxilla,[1–3] whereas others have stated that MIs inserted in the maxilla have attained higher success rates.[4] Similar conflicts have been reported with respect to gender,[2, 5–7] age,[6–8] MI length,[8, 9] and MI diameter.[4, 10]

To identify patterns among the different results of these studies, a meta−analysis can be applied. Meta−analyses have been performed to combine the outcomes of multiple studies into a single quantitative estimate, though statistical heterogeneity remains inevitable because of clinical and methodological differences among studies. For example, Dalessandri et al.[11] investigated factors that influence the success rates of temporary skeletal anchorage devices. They reported that treatment effects based on the patient's gender, age, and insertion site were heterogeneous between the studies and that this heterogeneity made it difficult to form conclusions. Since the results of a meta−analysis obtained by combining such heterogeneous effects are thus prone to be erroneous, it is necessary to specify a procedure to identify or eliminate the source of heterogeneity, when evaluating the outcomes obtained from diverse studies. En-masse retraction, canine retraction, and intrusion of posterior teeth are the 3 most common utilizations of the miniscrew. Clinically, placements in the buccal regions are easier and less variable, and more consistent. Since the inconsistent success rates become a reason of heterogeneity, a meta-analysis was performed on the success rates of buccal area only.

The aim of this study was to systematically review previous studies that addressed prognostic factors affecting the success of MIs inserted into the buccal posterior region, and to assess the combined odds ratio (OR) of the success of MI with respect to factors such as gender, age, jaw (mandible or maxilla), MI length, and MI diameter via a subgroup meta−analysis based on the study design, study quality, and sample size of MI.

**MATERIALS & METHODS**

Procedures for the meta–analysis complied with the Cochrane Handbook for Systematic Reviews of Interventions and the PRISMA statement.[12] Medline (PubMed), Scopus, and Web of Science electronic searches were conducted to obtain articles in English that were limited to clinical human studies published from January 2003 to March 2015, using the following search terms: factor(s), screw(s), implant(s), anchorage, success, stability, miniscrew(s), microimplant(s), and microscrew(s) (Appendix1). There has been no reported evidence that language restrictions imposed a bias in systematic review based meta-analyses. [13,14] In addition, manual searches of the reference list of electronically detected articles were performed, and a grey literature search was carried out using Google Scholar.

**Selection of Study**

The search and selection of articles were performed by three independent researchers (SH, EK, BK). Based on the candidate articles, two researchers (SH, EK) made a preliminary list of articles for the meta-analysis; Cohen's kappa was 0.88, indicating almost perfect agreement. Discordance in article selection was resolved by debate and consultation with another author (BK). As a result, 17 articles were subsequently selected.

**Inclusion Criteria**

The outcome measure of interest was the success of MI. This value was then converted into a dichotomous value: 0 for loosened MIs, and 1 for unloosened MIs. Five confounding factors were divided into two categories: patient factors and MI factors. The patient factors were age ($< 20$ years vs. $\geq 20$ years), gender (male vs. female), and jaw (maxilla vs. mandible), and the MI factors were length ($< 8$ mm vs. $\geq 8$ mm) and diameter ($\leq 1.4$ mm vs. $> 1.4$ mm). The following criteria were used to select appropriate articles: 1) studies on the stability of screws and implants, both used as orthodontic anchorage, 2) human clinical studies, 3) prospective and retrospective studies, 4) studies on MIs inserted into the posterior buccal region, and 5) studies that reported success rates of MIs or explicitly included information enabling a computation of the success rate regarding any of the five confounding factors.

**Exclusion Criteria**

Exclusion criteria included the following: articles related to 1) systematic review or meta-analysis, 2) patient's satisfaction, 3) orthognathic surgery, 4) radiographic evaluation, 5) microbiology, 6) case report, 7) in vitro, 8) literature review, 9) articles written by the same author, 10) studies based on the five confounding factors through having different dichotomizations: age factor with division of ($< 18$ years) vs. ($\geq 18$ years), and 11) studies on MIs inserted into retromolar pad, lingual side, or palatal side. Since one of the meta−analysis assumptions was independence between studies, studies from specific authors were included only once and studies from coauthors of the chosen studies were also excluded.

**Data Extraction**

The following information was extracted from the 17 included studies: author's name, publication year, study design, type of temporary anchorage device (TAD), diameter and length of MIs, number (No.) of patients, mean age, number of TADs, number of successes, success rate, definition of success and failure (Table 1). The ORs of success of MIs with respect to these factors were directly calculated, using the number of stable MIs and the number of inserted MIs, for each category. The ORs calculated from the raw data were double-checked by two authors (HK, HL). The success of MIs was defined as the absence of clinically detectable mobility when the orthodontic force was sustained during the predetermined period.

**Quality Assessment**

The quality of studies was assessed for non-randomized studies and then quantified using the 9-star Newcastle-Ottawa Scale (NOS)[15] recommended by the Cochrane collaboration. The study quality was evaluated as low (0-3 points), medium (4-6 points), and high (7-9 points). The NOS established eight criteria to evaluate the quality of the included studies, based on the three categories of subject selection, comparability between groups, and measurement of outcomes. Any disagreements were resolved by consensus (Table 2).

**Meta-analysis**

The outcome of the meta-analysis was expressed as a combined odds ratio (OR). The OR in each study was defined as the ratio of the odds (Success/Failure) of MIs in two categories: patient factors and MI factors. In the jaw, an OR greater than 1 indicates that the success of MIs more likely occurred in the maxilla (Jaw = 1), whereas an OR less than 1 indicates that the success more likely occurred in the mandible (Jaw = 0). Heterogeneity was tested by the Cochran Q and $I^2$ statistic. If the p value of the Q test is greater than 0.10, there is no significant heterogeneity. It was tentatively suggested that low heterogeneity may be associated with values of $I^2$ that are less than 30% and substantial heterogeneity with values of $I^2$ that are more than 50%; if there is significant heterogeneity, additional subgroup and sensitivity analyses were performed and a qualitative review of the study was conducted until the causes of the heterogeneity were clearly identified.

**Subgroup meta-analysis**

The studies included in the meta-analysis differed in study design, participant characteristics, and treatment goals. Variability among these studies in a systematic review may be taken as heterogeneity. To investigate the source of heterogeneity, a subgroup analysis can be used to answer specific questions about particular groups of patients, types of intervention or types of study. In this study, subgroups were created based on the study design (retrospective study vs. prospective study), study quality (medium vs. high), and sample size of MI (< 200 vs. ≥ 200). A meta-analysis was performed for each subgroup, and the results were reported separately.

**Sensitivity Analysis**

A sensitivity analysis is a collective method for verifying the robustness of results. It was performed to assess the impact of each study on the combined effect size. The meta-analysis was repeatedly performed as follows: a meta-analysis included all studies but the first one, the next meta-analysis included all but the second one, and this procedure was continued until every study was excluded once. If the statistical significance of the result was influenced by the removal of a single study, the removed study was reviewed again to confirm the source of the heterogeneity.

**Publication Bias**

Funnel plots have been widely used to detect the potential publication bias of studies in a meta-analysis. However, because the visual interpretation of funnel plots largely depends on the subjective impression of the observer, [16] Begg's rank correlation test [17] and Egger's linear regression test [18] were used as more objective tests to detect publication bias in this meta-analysis. Significant results ($p < 0.05$) suggest publication bias. Bias-corrected estimates were calculated using the trim and fill method, which accounts for unpublished data by imputing missing studies to yield an unbiased estimate of the effect size.

Meta-analyses and sensitivity analyses were performed using R-studio (v0.96.315, R studio Inc., USA) and Comprehensive Meta-Analysis Software (v2.0, Biostat, USA).

**RESULTS**

The preliminary electronic search found 2707 relevant articles. Articles that were not in English and not human clinical studies were excluded. Among the remaining 1696 articles, additional articles were excluded based on the aforementioned exclusion criteria. **After removing the duplicate publications,** the final 286 articles were then manually reviewed to determine whether they provided information facilitating the computation of MI success rates with respect to any of the five confounding factors. Finally, 17 articles satisfying all of the inclusion criteria were selected (Figure 1); the list of the included studies is shown in Table 1.

To investigate differences in the MI success rates with regards to the inserted jaw, 14 studies were used; 3 studies were excluded, since they did not provide success rates specifying the jaw of insertion. Admitting homogeneity ($P_{het}$(within) > 0.1) demonstrated by the 14 studies, the combined OR of 2.32 (95% CI = 1.81 − 4.08) indicated that MIs inserted in the maxilla had a 2.32 times significantly higher success rate than in the mandible. Each subgroup based on the study design, study quality, and sample size acquired homogeneity ($I^2$ < 25% and $p_{het}$(within) > 0.1). The combined OR of the prospective studies (OR = 3.67, 95% CI = 2.10−6.44) was higher than that of the retrospective studies (OR = 2.10, 95% CI = 1.60−2.74), and treatment effects differed between subgroups ($p_{het}$(between) = 0.077 < 0.1). The combined OR of studies having a high quality (OR = 2.18, 95% CI = 1.24 − 3.85) was lower than that of studies having a medium quality (OR = 2.36, 95% CI = 1.81 − 3.08), though treatment effects did not differ between the subgroups ($p_{het}$(between) > 0.1). The subgroup of 5 studies having a sample size not less than 200 MIs had a combined OR of 1.95 (95% CI = 1.42 − 2.68). Another subgroup of 9 studies having a sample size less than 200 MIs had a combined OR of 2.96 (95% CI = 2.04 − 4.29). Consequently, the subgroup meta-analysis revealed significantly higher success rates in the maxilla compared with the mandible. And the sensitivity analysis showed that none of the studies significantly changed the overall results of the subgroup analysis (Table 3, Figure 2).

In another meta-analysis investigating the difference in MI success rates according to gender, 13 of the 17 studies were considered; 4 studies were excluded since they did not provide the success rates according to gender. Admitting homogeneity ($P_{het}$(within) > 0.1) demonstrated by the 13 studies, the combined OR of 1.18 (95% CI = 0.92 − 1.51) indicated that there was no significant difference between the two genders. The subgroup of prospective studies showed homogeneity ($I^2$ = 3.73 and $P_{het}$(within) > 0.1), with a combined OR of 1.27 (95% CI = 0.63 − 2.54). The subgroup of retrospective studies also showed homogeneity ($I^2$ = 25.83 and $P_{het}$(within) = 0.206 > 0.1) and had a combined OR of 1.17 (95% CI = 0.89 − 1.52). Accordingly, there was no significant gender difference found with regards to MI success rates in either subgroup. Similar results were shown in the subgroup analyses based on study quality and sample size. Furthermore, the sensitivity analyses of the subgroups showed that none of the studies significantly changed the overall results (Table 3, Figure 3(a)).

For the meta-analysis investigating the difference in the MI success rates according to patient age, 6 studies were considered. Most excluded studies reported only the mean and standard deviation of age, and Moon et al.'s study[2] was excluded due to a different age dichotomization (over/under 18 years of age). Miyawaki et al.'s study[1] was excluded since the number of patients was used to calculate the frequency of the age categories instead of the number of TAD, unlike the other studies included in this meta-analysis. Based on homogeneity ($P_{het}$(within) > 0.1) demonstrated by the 6 studies, the combined OR of 1.59 (95% CI = 1.14 – 2.22) indicated that MIs inserted in patients of age 20 and over had 1.53 times significantly higher success rate than in patients under age 20. A subgroup of one prospective study and other subgroup of retrospective studies acquired homogeneity ($I^2$ =0.00 and $P_{het}$(within) > 0.1). The combined OR of the prospective study (OR = 3.23, 95% CI = 1.30–8.05) was higher than that of the retrospective studies (OR = 1.42, 95% CI = 1.03–1.96), but treatment effects did not differ between the subgroups ($P_{het}$(between) = 0.101 > 0.1). (Table 3, Figure 3(b)).

For the meta-analysis investigating the difference in MI success rates based on MI length, 4

studies including 628 MIs were used. Most individual studies considered in our meta‒analysis reported success rates based on MI length, though some excluded studies used only MIs having a long length($\geq$ 8mm)[1, 4, 5, 6, 10, 14, 17], and others did not provide success rates according to MI length restricted to the posterior region. Based on the homogeneity ($P_{het}$(within) > 0.1) demonstrated by the 4 studies, the combined OR of 0.46 (95% CI = 0.26 – 0.80) indicated that MIs with long length($\geq$ 8mm) had 2.17(=1/0.46) times significantly higher success rates than MIs with short length(< 8mm). While the combined OR of the prospective study (OR = 0.56, 95% CI = 0.19 – 1.64) showed that there was no significant differences between long length and short length, the combined ORs of the retrospective studies (OR = 0.42, 95% CI = 0.22 – 0.82) indicated that MIs with long length had significantly higher success rate than MIs with short length. Homogeneity was also obtained in all subgroups based on the study design, study quality, and sample size ($I^2$ < 25%). (Table 3, Figure 3(c)).

In the meta‒analysis evaluating the difference in MI success rates according to MI diameter, 4 studies were considered. Some excluded studies used only one type of MI diameter. Other studies included different diameters, but did not report success rates according to the diameter. Based on the homogeneity ($P_{het}$(within) > 0.1) demonstrated by the 4 studies, the combined OR of 0.62 (95% CI = 0.40 – 0.97) indicated that MIs with large diameter(> 1.4mm) had 1.61(=1/0.62) times significantly higher success rates than MIs with small diameter(≤1.4mm). While the combined OR of the retrospective studies (OR = 0.74, 95% CI = 0.45 – 1.22) indicated that there was no significant difference between large diameter and small diameter, the combined OR of the prospective study, Wiechmann et al.'s study[10], (OR = 0.34, 95% CI = 0.14 – 0.86) showed that MIs with large diameter had significantly higher success rates than MIs with small diameter. (Table 3, Figure 3(d)).

Publication bias was assessed for the five factors used in our study. No publication bias was found based on the Begg's and Eggar's tests (p > 0.05), except in the subgroup of retrospective studies regarding age. If publication bias was found, a bias−corrected estimate OR obtained by the trim−and−fill method was used as the final outcome. Note that though the OR values changed, the significances did not change (Table4).

**DISCUSSION**

The meta–analysis included both prospective and retrospective studies, and the OR was used as the effect size since the OR can be used for the three major study designs: cross–sectional, prospective, and retrospective studies.[19, 20] Meta analyses regarding 5 factors—age, gender, jaw, MI length, and MI diameter—were performed in Dalessandri et al.'s study.[11] In their study, gender, MI length, and MI diameter showed no significant differences, but MIs inserted in the maxilla had higher success rates than those in the mandible and MIs inserted in older (>20) people had higher success rates than those in younger (<20) people. While jaw (maxilla/mandible), gender, and age in the results of Dalessandri et al.'s study[11] were consistent with this study's results, those for the length and diameter were inconsistent. In Papageorgion et al.'s study,[21] age, gender, MI length, and MI diameter showed no significant differences. They reported MIs inserted in the maxilla had higher success rates than those in the mandible. Their results for the jaw and gender only were consistent with ours; however, those for age, length, and diameter were inconsistent. In Crismani et al's study[22], only a systematic review was performed, with no meta-analysis. They reported that screws under 8mm in length and 1.2 mm in diameter should be avoided, which was consistent with our results.

**Jaws**

MIs inserted into the anterior region and the palatal side that had serious effects on the outcome were excluded from our meta–analysis because greater root proximity in the anterior region [23] and significantly higher success rate on the palatal side [24] have been reported. Since the success rates of miniscrews placed in the mandibular lingual side was lower than for the buccal side, the success rate in the mandible was seen to be much lower than for the maxilla.[25] Moreover, since the retromolar area showed the highest success rate (100%) and this area was included in the mandible, the success rate in the mandible was higher than that of the maxilla.[26] Finally, due to the fact that these inconsistent sucess rates became a reason for the heterogeneity, a meta-analysis was performed on the success rates of the buccal area only. MIs inserted in the maxilla had higher success rates than those inserted in the mandible regardless of the study design. Furthermore, prospective studies showed higher success rates for the MIs inserted in the maxilla compared to retrospective studies. Since prospective studies usually have fewer potential sources of bias and confounding than retrospective studies, the prospective studies are deemed more reliable than retrospective studies. Higher success rates in the maxilla than in the mandible were reported due to thicker mandibular cortical bone than the maxillary bone and overheating of the mandibular bone during drilling and irritating during chewing [3,4]. While the insertion region in the maxilla is keratinized gingiva, there is a high possibility that the mandibular insertion region is free gingiva. For this reason, gingival inflammation is likely to occur more in the mandible compared to maxilla. Management of oral hygiene can help to prevent the miniscrew from loosening. Another recommended way to reduce the MI insertion failure in the mandibular buccal posterior region is the pilot drilling, which is different from the pre-drilling. During the pilot drilling, the MI is inserted into the notch with the fissure bur, using a hand driver.[27]

**Gender**

Since females tend to get more orthodontic treatments than males for esthetic reasons, the proportion of females was significantly larger in most studies considered in our meta–analysis. Although no heterogeneity among the 13 studies was detected, a subgroup analysis was performed using the criterion of the study design, study quality, and sample size of MI; each subgroup acquired homogeneity. Although males have a higher bone mineral density than females, the success rate of MI was not seen to be significantly different between the two genders. These findings were consistent with previous studies.[28]

**Age**

In this meta–analysis, significantly higher success rates were shown in patients of age 20 and over (age ≥ 20) compared to patients under age 20 (age < 20). But in studies that had a sample size of less than 200 MIs, there was no significant difference between the success rates of the MIs inserted in patients that had an age of 20 and over those who were under age 20. However, since these studies had a moderate heterogeneity ($I^2 = 32.54\%$) – a few of the studies had a different success rate – these results should be interpreted cautiously. Chen et al. (2007)[29] reported that adolescents had a higher chance of loosening than adults, as a thin cortical bone and low bone density were linked to the increased failure of MIs. This result was consistent with ours. Removable appliances or extraoral appliances such as facemask may be suitable alternatives to miniscrew for adolescents. Females usually reach full physical development by age 15–17, whereas males typically complete puberty by age 18–19.[30] If the data in the previously published articles had been categorized by age group considering these peaks of physical growth, the results might enhance our results of the meta–analysis.

**Length and diameter**

The most common lengths in buccal area are 6mm and 8mm. Based on Crismani et al.'s study,[22] which was recommended that lengths under 8 mm be avoided, the division was made as <8mm vs. ≥8mm. As for diameter, ≤1.4mm vs. >1.4mm was selected because the common diameters in buccal region are 1.4mm and 1.6mm, Crismani et al.[22] recommended that miniscrew diameters under 1.2mm should be avoided, and Kuroda et al.[31] recommended that miniscrew diameters of 1.3mm and over be used. Even when the criteria of diameter is changed to ≤1.2mm vs. >1.2mm or ≤1.3mm vs. >1.3mm, the larger diameter showed a higher success rate than the smaller one. Note, however, that though a significant difference was found between the groups up to 1.4mm, from the groups of ≤1.5mm vs. >1.5mm, there was no significance observed. This result supports the results reported by Crismani et al.[22] and Kuroda et al.[31] Due to the fact that miniscrew retention depends on its mechanical interdigitation to the bone, the physical length and diameter of the screw plays an essential role in their placement. Compared to longer and larger screws, smaller and shorter screws have less mechanical interdigitation and can easily break off during implantation. It is thus important to select an appropriate size when miniscrews are placed in the interradicular space to provide orthodontic anchorage, and it is also essential to prevent the screw's proximity to roots.[31] In terms of clinical applications, if we know the success rate of each type of screw (length/diameter) as well as its region (maxilla/mandible), we will be able to reduce the failure rate based on the findings of this meta-analysis. More well−designed studies will be needed to make a firm conclusion, and authors should choose appropriate studies not to compare apples and oranges.

**CONCLUSION**

- The success rates of the MIs inserted in the maxilla were higher than those in the mandible. Compared to retrospective studies, prospective studies showed higher success rates in the MIs inserted in the maxilla.

- Significantly higher MI success rates were found in patients that had an age of 20 and over, MIs with long lengths (≥8mm), and large diameters (>1.4mm).

- There was no significant difference found between the success rates of the MIs inserted in males and females. When a treatment plan is made, these risk factors should be taken into account.

**ACKNOWLEDGMENTS**

This research was supported by the Basic Science Research Program of the National Research Foundation of Korea (NRF), funded by the Ministry of Education (2011−0012875).

The authors declare no potential conflicts of interest with respect to the authorship and/or publication of this article.

**FIGURE LEGENDS**

**Figure 1.** PRISMA flow diagram of search strategy

**Figure 2.** (a) Forest plot of jaw by subgroup analysis
(b) Sensitivity analysis of 11 studies in the retrospective study
(c) Sensitivity analysis of 3 studies in the prospective study

**Figure 3.** (a) Forest plot of gender by subgroup analysis
(b) Forest plot of age by subgroup analysis
(c) Forest plot of MI length by subgroup analysis
(d) Forest plot of MI diameter by subgroup analysis

(a) Forest plot of jaw (mandible/maxilla) by subgroup analysis

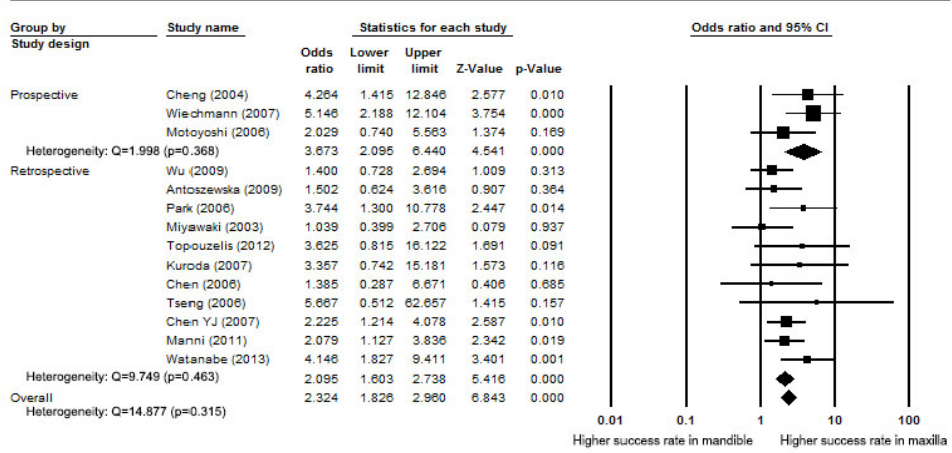

(b) Sensitivity analysis of prospective studies

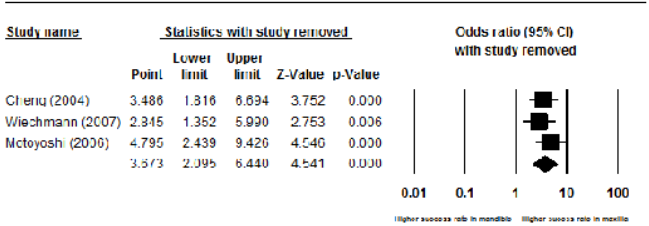

(c) Sensitivity analysis of retrospective studies

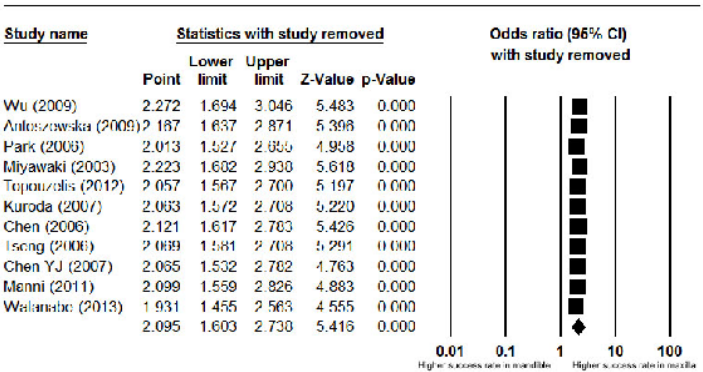

**Figure 2.** (a) Forest plot of ORs for success rates between mandible and maxilla. Homogeneous subgroup of prospective studies showed combined OR of 3.67 (95% CI=2.10-6.44)[*], and homogeneous subgroup of retrospective studies showed a combined OR of 2.10 (95% CI= 1.60-2.74)[*]. (b) In the sensitivity analysis for prospective studies regarding jaw, a significantly higher success rate in the maxilla was shown. (c) The sensitivity analysis of retrospective studies regarding jaw showed a higher success rate in the maxilla.

Fig. 1. (a) Forest plot of jaw by subgroup analysis
(b) Sensitivity analysis of 11 studies in the retrospective study
(c) Sensitivity analysis of 3 studies in the prospective study

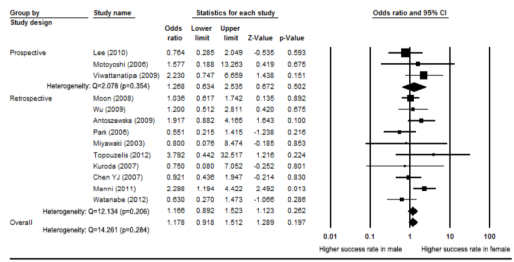
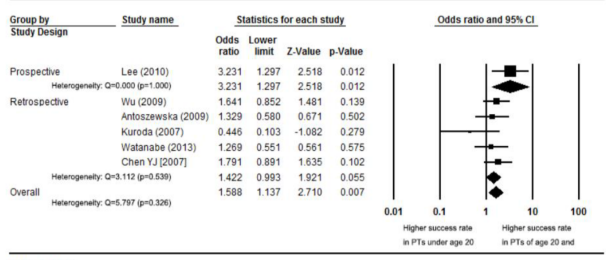
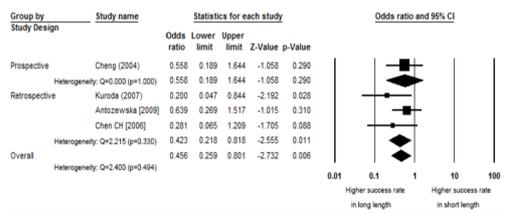
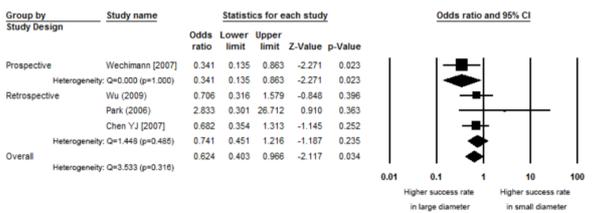

**Figure 3**. (a) Forest plot of ORs for success rates between males and females. The results in both subgroups showed no significant difference in success rates. (b) Forest plot of ORs for success rates between age groups. Homogeneous subgroups showed higher success rates in patients aged 20 and over (≥ 20) except subgroup of retrospective studies and studies with small samples. (c) Forest plot of ORs for success rates between long MIs (≥8mm) and short MIs (<8mm). No significant differences were found. (d) Forest plot of ORs for success rates between large diameter (>1.4 mm) and small diameter (≤1.4 mm) screws. No significant differences were found.

Fig. 2. (a) Forest plot of gender by subgroup analysis
(b) Forest plot of age by subgroup analysis
(c) Forest plot of MI length by subgroup analysis
(d) Forest plot of MI diameter by subgroup analysis

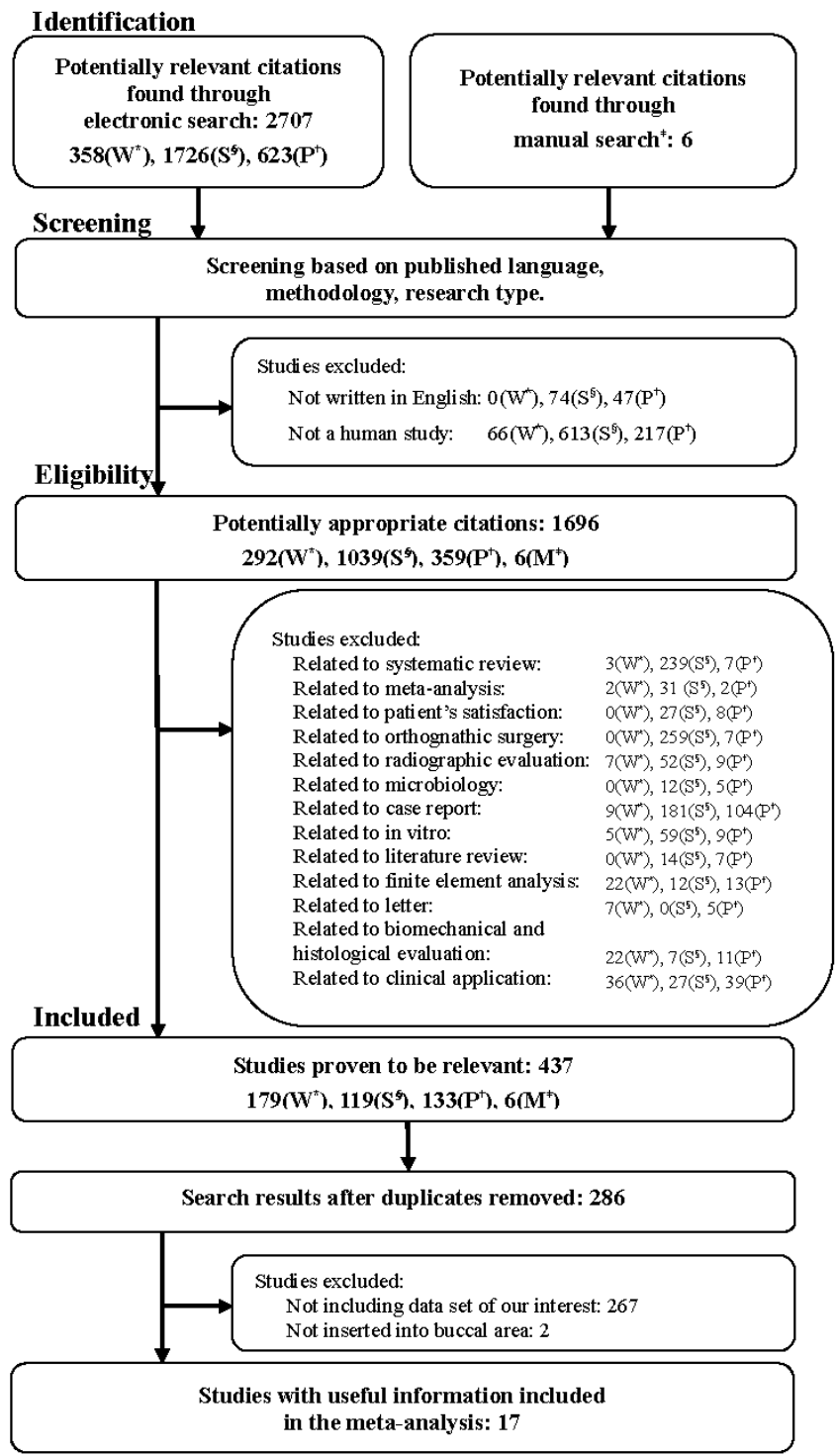

Figure 1. PRISMA flow diagram of search strategy
*W: Web of Science, §S: Scopus, †P: Pubmed, ‡M: Manual search

Fig. 3. PRISMA flow diagram of search strategy

**Table 2a.** Newcastle-Ottawa Scale (NOS) summary assessment of risk of bias for retrospective studies

| Study | Is the case definition adequate? | Representativeness of cases | Selection of controls | Definition of controls | Comparability of cases and controls on the basis of the design or analysis | Ascertainment of exposure | Same ascertainment method for cases and controls | Nonresponse rate | NOS score | Overall assessment |
|---|---|---|---|---|---|---|---|---|---|---|
| Moon et al[2] 2008 | 1 | 1 | 0 | 0 | 2 | 1 | 1 | 0 | 6 | medium |
| Wu et al[7] 2009 | 1 | 1 | 0 | 0 | 0 | 1 | 1 | 0 | 4 | medium |
| Chen YJ et al[33] 2007 | 1 | 1 | 0 | 0 | 2 | 1 | 1 | 0 | 6 | medium |
| Antoszewska et al[27] 2009 | 1 | 1 | 0 | 1 | 0 | 1 | 1 | 0 | 5 | medium |
| Manni et al[30] 2011 | 1 | 0 | 0 | 1 | 1 | 1 | 1 | 0 | 5 | medium |
| Park et al[4] 2006 | 1 | 0 | 0 | 1 | 1 | 1 | 1 | 0 | 5 | medium |
| Watanabe et al[34] 2013 | 1 | 1 | 0 | 0 | 0 | 1 | 1 | 0 | 4 | medium |
| Miyawaki et al[1] 2003 | 1 | 1 | 0 | 0 | 2 | 1 | 1 | 0 | 6 | medium |
| Topouzelis et al[29] 2012 | 1 | 0 | 0 | 1 | 1 | 1 | 1 | 0 | 5 | medium |
| Kuroda et al[8] 2007 | 1 | 1 | 0 | 0 | 0 | 1 | 1 | 0 | 4 | medium |
| Chen CH et al[28] 2006 | 1 | 1 | 0 | 1 | 0 | 1 | 1 | 0 | 5 | medium |
| Tseng et al[32] 2006 | 1 | 0 | 0 | 1 | 0 | 1 | 1 | 0 | 4 | medium |

**Table 2b.** Newcastle-Ottawa Scale (NOS) summary assessment of risk of prospective studies

| Study | Representativeness of the exposed cohort | Selection of the nonexposed cohort | Ascertainment of exposure | Outcome of interest not present at start of study | Comparability | Assessment of outcome | Adequacy of duration of follow-up | Adequacy of completeness of follow-up | Total score | Overall assessment |
|---|---|---|---|---|---|---|---|---|---|---|
| Lee et al[6] 2010 | 0 | 0 | 1 | 1 | 1 | 0 | 1 | 1 | 5 | medium |
| Cheng et al[31] 2004 | 1 | 0 | 1 | 1 | 2 | 1 | 1 | 1 | 8 | high |
| Wiechmann et al[10] 2007 | 0 | 0 | 1 | 0 | 0 | 1 | 1 | 1 | 4 | medium |
| Motoyoshi et al[3] 2006 | 0 | 0 | 1 | 1 | 2 | 0 | 1 | 1 | 6 | medium |
| Viwattanatipa et al[9] 2009 | 0 | 0 | 1 | 1 | 2 | 1 | 1 | 1 | 7 | high |

**Table 3.** Subgroup analyses based on sample size of Miniscrew Implant (MI), study type, and study quality with respect to five confounding factors.

| | OR(95% CI) | n | I²(%) | P$_{het}$(within) | P$_{het}$(between) |
|---|---|---|---|---|---|
| **Jaw** | | | | | |
| Study design | | | | | |
|   Prospective | 3.67(2.10, 6.44) | 3 | 0.00 | 0.368 | |
|   Retrospective | 2.10(1.60, 2.74) | 11 | 0.00 | 0.463 | 0.077 |
| Study quality | | | | | |
|   High | 2.18(1.24, 3.85) | 3 | 6.17 | 0.344 | |
|   Medium | 2.36(1.81, 3.08) | 11 | 21.16 | 0.242 | 0.775 |
| Sample size | | | | | |
|   ≥200 | 1.95(1.42, 2.68) | 5 | 0.00 | 0.556 | |
|   <200 | 2.96(2.04, 4.29) | 9 | 11.94 | 0.335 | 0.119 |
| Overall | 2.32(1.83, 2.96) | 14 | 12.62 | 0.315 | |
| **Gender** | | | | | |
| Study design | | | | | |
|   Prospective | 1.27(0.63, 2.54) | 3 | 3.73 | 0.354 | |
|   Retrospective | 1.17(0.89, 1.52) | 10 | 25.83 | 0.206 | 0.805 |
| Study quality | | | | | |
|   High | 2.23(0.75, 6.66) | 1 | 0.00 | 1.000 | |
|   Medium | 1.14(0.88, 1.47) | 12 | 14.61 | 0.301 | 0.238 |
| Sample size | | | | | |
|   ≥200 | 1.19 (0.91, 1.58) | 7 | 36.09 | 0.153 | |
|   <200 | 1.11(0.63, 1.97) | 6 | 0.00 | 0.438 | 0.872 |
| Overall | 1.18(0.92, 1.51) | 13 | 15.85 | 0.284 | |
| **Age** | | | | | |
| Study design | | | | | |
|   Prospective | 3.23(1.30, 8.05) | 1 | 0.00 | 1.000 | |
|   Retrospective | 1.42 (0.99, 2.04) | 5 | 0.00 | 0.539 | 0.101 |
| Study quality | | | | | |
|   High | | | | | |
|   Medium | 1.59(1.14, 2.22) | 6 | 13.75 | 0.326 | 1.000 |
| Sample size | | | | | |
|   ≥200 | 1.81(1.24, 2.64) | 4 | 0.00 | 0.538 | |
|   <200 | 0.98(0.48, 2.03) | 2 | 32.54 | 0.223 | 0.230 |
| Overall | 1.59(1.14, 2.22) | 6 | 13.75 | 0.326 | |
| **Length** | | | | | |
| Study design | | | | | |
|   Retrospective | 0.42(0.22, 0.82) | 3 | 9.72 | 0.330 | |
|   Prospective | 0.56(0.19, 1.64) | 1 | 0.00 | 1.000 | 0.641 |
| Study quality | | | | | |
|   Medium | 0.42(0.22, 0.82) | 3 | 9.72 | 0.330 | |
|   High | 0.56(0.19, 1.64) | 1 | 0.00 | 1.000 | 0.641 |
| Sample size | | | | | |
|   ≥200 | 0.64(0.27, 1.52) | 1 | 0.00 | 1.000 | |
|   <200 | 0.36(0.17, 0.75) | 3 | 0.00 | 0.501 | 0.313 |
| Overall | 0.46(0.26, 0.80) | 4 | 0.00 | 0.494 | |
| **Diameter** | | | | | |
| Study design | | | | | |
|   Retrospective | 0.74(0.45,1.22) | 3 | 0.00 | 0.485 | |
|   Prospective | 0.34(0.14,0.86) | 1 | 0.00 | 1.000 | 0.149 |
| Study quality | | | | | |
|   Medium | 0.62(0.40,0.97) | 4 | 15.10 | 0.316 | 1.000 |
|   High | | | | | |
| Sample size | | | | | |
|   ≥200 | 0.74(0.45,1.22) | 3 | 0.00 | 0.485 | |
|   <200 | 0.34(0.14,0.86) | 1 | 0.00 | 1.000 | 0.149 |
| Overall | 0.62(0.40,0.97) | 4 | 15.10 | 0.316 | |

OR: Odds Ratio, CI: Confidence Interval
P$_{het}$(within): heterogeneity within subgroups, P$_{het}$(between): heterogeneity between subgroups.
I²>50: substantial heterogeneity

**Appendix 1.** Search strategy for the electronic databases used in this meta-analysis

| Databases of published articles | Search strategy used |
|---|---|
| **Pubmed** http://www.ncbi.nlm.nih.gov/pubmed | ((((((((factor[ti]) OR factors[ti]) OR success[ti]) OR failure[ti]) OR orthodontic[ti]) OR anchorage[ti]) OR stability[ti]) AND ((((((((((((((((miniimplant[ti]) OR miniimplants[ti]) OR mini implant[ti]) OR mini implants[ti]) OR mini-implant[ti]) OR mini-implants[ti]) OR miniscrew[ti]) OR miniscrews[ti]) OR mini screw[ti]) OR mini screws[ti]) OR mini-screw[ti]) OR mini-screws[ti]) OR microscrew[ti]) OR microscrews[ti]) OR micro screw[ti]) OR micro screws[ti]) OR micro-screw[ti]) OR micro-screws[ti]) OR microimplant[ti]) OR microimplants[ti]) OR micro implant[ti]) OR micro implants[ti]) OR micro-implant[ti]) OR micro-implants[ti]) NOT (systematic review[ti]) NOT (meta-analysis) NOT (patient satisfaction) NOT (orthognathic surgery) NOT (radiographic evaluation) NOT (microbiology) NOT (in vitro) NOT (case reports) NOT (case report) NOT (literature review) NOT (literature reviews)) |
| **Scopus** http://www.scopus.com | ALL(factor OR success OR failure OR orthodontic OR anchorage OR stability) AND ALL(miniimplant* OR "mini implant*" OR "mini-implant*" OR miniscrew* OR "mini screw*" OR "mini-screw*" OR microscrew* OR "micro screw*" OR "micro-screw*" OR microimplant* OR "micro implant*" OR "micro-implant*") AND PUBYEAR > 2002 AND PUBYEAR < 2015 AND LANGUAGE(english) AND NOT "systematic review" AND NOT "meta analysis" AND NOT "patient satisfaction" AND NOT "orthognathic surgery" AND NOT "radiographic evaluation" AND NOT microbiology AND NOT "in vitro" AND NOT "case report" AND NOT "literature review" AND ( LIMIT-TO(DOCTYPE,"ar") ) AND ( LIMIT-TO(SUBJAREA,"DENT") OR LIMIT-TO(SUBJAREA,"MULT") ) |
| **Web of Science** http://www.webofknowledge.com | TI=((factor OR success OR failure OR orthodontic OR anchorage OR stability) AND ((miniimplant) OR (mini implant) OR (mini-implant) OR (miniscrew) OR (mini screw) OR (mini-screw) OR (microscrew) OR (micro screw) OR (micro-screw) OR (microimplant) OR (micro implant) OR (micro-implant)) NOT (systematic review) NOT (meta-analysis) NOT (patient satisfaction) NOT (orthognathic surgery) NOT (radiographic evaluation) NOT (microbiology) NOT (in vitro) NOT (case report) NOT (literature review)) AND SU=Dentistry NOT TS=(Animal OR Beagle OR Rabbit OR Rat OR pig) |

**Table 1**. General characteristics of the studies included in the meta-analysis

| | Author Year | Study Design Type of TAD | Diameter (mm) | Length (mm) | N. of patients | Mean age | N. of TAD | N. of Success | Success rate | Definition of Success | Definition of Failure |
|---|---|---|---|---|---|---|---|---|---|---|---|
| 1 | Lee et al[6] 2010 | Prospective Miniscrew | 1.8 | 8.5 | 141 | 27 | 260 | 238 | 91.54 | NA | NA |
| 2 | Cheng et al[33] 2004 | Prospective Miniscrew | 2.0 | 5/7/9/11/13/15 | 44 | 29±8.9 | 140 | 125 | 89.29 | Absence of inflammation and clinically detectable mobility of sustaining the anchorage function | NA |
| 3 | Wiechmann et al[10] 2007 | Prospective Microimplants | 1.1/1.6 | 5/6/7/8/10 | 49 | 26.9±8.9 | 133 | 102 | 76.69 | Absence of inflammation and clinically detectable mobility, capability of sustaining the anchorage function | NA |
| 4 | Motoyoshi et al[3] 2006 | Prospective Microimplants | 1.6 | 8 | 41 | 24.9±6.5 | 124 | 106 | 85.48 | Endured orthodontic force for 6 months or more | Loosened before 6 months |
| 5 | Viwattanatipa et al[9] 2009 | Prospective Miniscrew | 1.2 | 8/10/12 | 49 | 23.2 | 97 | 65 | 89.04 | NA | Remarkable mobility dislodgement, infection |
| 6 | Moon et al[2] 2008 | Retrospective Miniscrew | 1.6 | 8 | 209 | 20.3 | 480 | 402 | 83.75 | Did not show any mobility after first 8 month. | NA |
| 7 | Wu et al[7] 2009 | Retrospective Mini-implants | 1.1-1.5/1.7/2.0 | 7/8/10/11/12/13/14/15 | 166 | 26.5±8.9 | 414 | 372 | 89.86 | NA | Loosened within 6 month or fractured during insertion |
| 8 | Chen YJ et al[28] 2007 | Retrospective Miniscrew/microscrew | 2.0/1.2 | 5-21/6-10 | 129 | 24.5±7.1 | 359 | 306 | 85.20 | NA | Loosened during treatment |
| 9 | Antoszewska et al[36] 2009 | Retrospective Miniscrew | 1.2-1.3/1.2-1.6 | 6/8 | 141 | NA | 350 | 327 | 93.43 | Absence of inflammation and clinically detectable mobility, capability of sustaining the anchorage function | Severe clinical mobility, loss of MI while checking its mobility with the cotton tweezers less than 8 month |
| 10 | Manni et al[38] 2011 | Retrospective Miniscrew | 1.5/1.3 | 9/11 | 132 | 23.2 | 300 | 243 | 81.00 | Absence of inflammation and clinically signs of loosening | Clinical signs of inflammation, instability of the miniscrew |
| 11 | Park et al[4] 2006 | Retrospective Miniscrew | 1.2/2.0 | 4/5/6/7/8/10/12/14/15 | 87 | 15.5±8.3 | 227 | 208 | 91.63 | Maintained to the end of treatment or to intentional removal | Loosened during treatment |
| 12 | Watanabe et al[35] 2013 | Retrospective Miniscrew | 1.4 | 5/6/8 | 107 | 21.0 | 190 | 162 | 85.26 | NA | Showed mobility or failed within 1 year of placement |
| 13 | Miyawaki et al[1] 2003 | Retrospective miniscrew | 1.0/1.5/2.3 | 6/11/14 | 51 | 21.8±7.8 | 134 | 104 | 77.61 | If orthodontic force could be applied for 1 year | NA |
| 14 | Topouzelis et al[32] 2012 | Retrospective Miniscrew | 1.2/1.4 | 8/10 | 34 | 27.2±7.3 | 82 | 74 | 90.24 | No inflammation or clinically detectable mobility present nor any dental root damage | Infection, dislodgement or remarkable mobility that could not sustain orthodontic force |
| 15 | Kuroda et al[8] 2007 | Retrospective Miniscrew | 1.3/2.0/2.3 | 6/7/8/10/11/12 | 75 | 21.8±8.2 | 79 | 70 | 88.61 | Applied to the skeletal anchorage for 1 year | NA |
| 16 | Chen CH et al[37] 2006 | Retrospective Microimplants | 1.2 | 6/8 | 29 | 29.8 | 59 | 50 | 84.75 | Keep the retention of anchors, absence of inflammation, no dental root damage | Severe mobility, persistence of infection of inflammation |
| 17 | Tseng et al[34] 2006 | Retrospective Mini-implants | 2.0 | 8/10/12/14 | 25 | 29.9 | 45 | 41 | 91.11 | Resist orthodontic force until completion of the orthodontic treatment, no inflammation or infection | Severe mobility, persistence of infection of inflammation |

MI: Miniscrew Implant, TAD: temporary anchorage device